\documentclass[letterpaper,titlepage,10pt]{article}
\usepackage{setspace} 
\usepackage[colorlinks=true,urlcolor=blue,citecolor=magenta]{hyperref}
\usepackage{xcolor}
\usepackage{amssymb,amsmath,amsfonts}
\usepackage{epsfig}
\usepackage{epsfig}
\usepackage{graphicx}
\usepackage{epstopdf}
\usepackage[round]{natbib}
\usepackage{caption}
\usepackage{subcaption}
\usepackage{amscd}
\usepackage{amsthm}
\usepackage{csquotes}
\usepackage{latexsym}
\usepackage{float}
\usepackage{mathtools}
\usepackage[stable]{footmisc}
\usepackage{amsbsy}
\usepackage{bbm}
\usepackage[english]{babel}
\usepackage{psfrag}
\usepackage{tabularx}
\usepackage{csquotes}
\usepackage{colortbl}
\usepackage{braket}
\usepackage{subfloat}
\usepackage{lipsum}                     
\usepackage{xargs}                      

\usepackage[colorinlistoftodos,prependcaption,textsize=tiny]{todonotes}
\newcommand{\beq}{\begin{eqnarray}}
\newcommand{\eeq}{\end{eqnarray}}

\newcommandx{\unsure}[2][1=]{\todo[linecolor=red,backgroundcolor=red!25,bordercolor=red,#1]{#2}}
\newcommandx{\change}[2][1=]{\todo[linecolor=blue,backgroundcolor=blue!25,bordercolor=blue,#1]{#2}}
\newcommandx{\info}[2][1=]{\todo[linecolor=OliveGreen,backgroundcolor=OliveGreen!25,bordercolor=OliveGreen,#1]{#2}}
\newcommandx{\improvement}[2][1=]{\todo[linecolor=Plum,backgroundcolor=Plum!25,bordercolor=Plum,#1]{#2}}
\newcommandx{\thiswillnotshow}[2][1=]{\todo[disable,#1]{#2}}

\usepackage{enumitem}
\setlistdepth{9}
\renewlist{enumerate}{enumerate}{9}
\setlist[enumerate]{leftmargin=*, labelsep=0.5em, itemsep=0.2em, topsep=0.4em}
\setlist[enumerate,1]{label=\arabic*.}
\setlist[enumerate,2]{label*=\arabic*.}
\setlist[enumerate,3]{label*=\arabic*.}
\setlist[enumerate,4]{label*=\arabic*.}
\setlist[enumerate,5]{label*=\arabic*.}
\setlist[enumerate,6]{label*=\arabic*.}
\setlist[enumerate,7]{label*=\arabic*.}
\setlist[enumerate,8]{label*=\arabic*.}
\setlist[enumerate,9]{label*=\arabic*.}

\begin{document}
\centerline{ \bf \Large Tractatus Quanticus}

\vskip .3cm
\centerline{\large {{\em Niccolò Covoni $^{1,2}$, Carlo Rovelli $^{3,4}$}}}
\vskip .3cm
\begin{center}
\sl $^1$ DISPeA, University of Urbino, 
\sl Via Timoteo Viti n.10, 61029 Urbino (PU) Italy.\\
\sl $^2$ USI, University of Svizzera Italiana, 
 Via Buffi n.13, 6900 Lugano, Switzerland.\\ 
\sl $^3$Aix-Marseille University, Universit\'e de Toulon, 
 CPT-CNRS, F-13288 Marseille, France\\
\sl $^4$ The Rotman Institute of Philosophy, 
 1151 Richmond St.~N London  N6A5B7, Canada.
\end{center}
\vskip-0.1cm
\centerline{\small\tt n.covoni@campus.uniurb.it, rovelli.carlo@gmail.com}

\vskip .6cm \centerline{\bf Abstract} \vskip 0.2cm 

{\em \noindent This text will perhaps only be understood by those who have themselves already thought the thoughts which are expressed in it, or similar thoughts. Its meaning could be summed up somewhat as follows: whatever can be said at all, can be said clearly by any speaker who is part of the world that is spoken about, and who speaks from its perspective.\\
\noindent Of course others far better than us have already been around here. But we feel some details have been left out, which quantum phenomena have brought into sharp light, and these bring further clarity. \\ Whoever understands us recognizes again the following propositions as senseless. After climbing through these, over them so to speak, will kick away the ladder, once more.  \\ Thanks to this futility, we feel shamelessly entitled to parrot a Master. 
\noindent  Inspired not by his preaching but rather by his example (and not in small part by its later reflections), we have written all this, as we actually think that not to remain silent, even whereof one cannot speak, is not such a bad idea after all, maybe only, as we learned from him, to keep the warning alive that there are questions that have no meaning. \\ What we hold onto is not knowledge of an absolute order, but the fleeting grasp of a perspective on perspectives. Hence this is at most just a manifesto for a picture of the world suggested by quantum mechanics. Each proposition  attempts to show a possible way of formulating questions about nature in a manner coherent with what we have recently learned about it. A manner aware of its own partiality.  Good philosophy, seems to us, dismantles idols, rather than creating them, and, like a gentle friend, reminds us with indirect hints, over and over again, that the answer to the riddle is that there is no riddle.}

\vskip 0.4cm


\vspace{.6cm}


\section*{Synthesis}
\begin{enumerate}
	\item The world is everything that is the case from some perspective. 
	
	\item My knowledge about the world is a particular kind of perspective. 
	
	\item Perspectives themselves are facts, when considered from another perspective. 
	
	\item Facts can be expressed as values of variables, which are ways systems interact. 
	
	\item Perspectives are transparent to each other, because they are facts.
	
	\item There is nothing wrong with circularity. 
	
	\item What we do not have information about, we must pass over in silence. 
\end{enumerate}


\vspace{1cm}

\section*{Short}

\begin{enumerate}
  \item \textbf{The world is everything that is the case from some perspective.}
       \begin{enumerate}
    \item Everything that is the case, is the case from some perspective.
    \item Something that is the case from a perspective is a fact, from (or in) a perspective.\label{fact}
    \item Our perspectives are special cases of the physical notion of perspective.
      \item Perspectives are not about minds. They are about physics, like reference systems in relativistic physics are about physics, not about minds.
      \item No perspective stands out above the rest.
  \end{enumerate}
  \item \textbf{My knowledge about the world is a particular kind of perspective.\footnote{Here, ``particular kind" denotes a concrete instantiation of what is assumed to be universal. The world is constituted by particular perspectives, whereas a universal is merely the sum of these particulars. A whole that can always be extended by adding a further particular. }}
  \begin{enumerate}
    \item My knowledge\footnote{ We use the term ``knowledge" in the \cite{WittgensteinPI}'s sense \S116, where knowing is embedded in practices and does not imply access to a total view or a metaphysical picture.} of the world is the information I have or can have about the world.
    \item Information has degrees of reliability.\footnote{Cfr. \cite{De1989-FINPAC-2}.}
    \item A fact in my perspective is knowledge about which I (reasonably) estimate no uncertainty.
    \item Degrees of reliability assigned to facts evolve; perspective can change accordingly.
      \item To have uncertain information about a fact means to have the information that by gathering more information the fact will be ascertained with a known likelihood.
    \item If I reliably expect that one out of \(N\) mutually exclusive alternatives \(a\) will be the case, then I can talk about a set of alternative possible facts.
    \item Facts in a perspective can be represented in logical spaces. They can be actual, or have probability.
    \item My knowledge about the world does not create the world; it is determined by the world and it refers to it.
  \end{enumerate}
  \item \textbf{Perspectives themselves are facts, when considered from another perspective.}\label{3}
  \begin{enumerate}
    \item A perspective is embodied in a set of facts, when considered from any other perspective (including itself).
    \item If knowing \(b\) determines which \(a\) is a fact, then \(a\) is fact with respect to \(b\), or in the perspective of \(b\).
    \item If \(p(a,b)\) is the joint probability for two sets of possible facts \(A=\{a\}\) and \(B=\{b\},\) then \(b\) embodies a perspective in which the probabilities of \(a\) are \(p_b(a) = p(a|b) = p(a,b)/p(b).\)
    \item To know \textit{what} \(B\) knows about \(A\) is not the same as knowing \textit{that} \(B\) knows about \(A.\)
    \item The mutual information \(I_{A:B}\) quantifies the information on one variable that can be obtained from the other.
 \item When the relative information \(I_{A|B}\)  is maximal, knowing \(b\) determines which \(a\) is a fact, and \(a\) is a fact relative to \(b\).\footnote{\cite{RelativeinfoRovelli}.}
    \item All things stand only in relation to a perspective, and with a shift of perspective, what once appeared as fact may be absent or incompatible with other facts from another perspective.
    \item A subject is a (case of a) perspective and defines a world.
    \item From a perspective distinct from mine, I, a subject, am a physical process, namely a set of facts.
    \item Knowledge is physically underpinned by relative information between variables of two systems: the subject and the object of knowledge. \label{knowledge}
    \item The world for me is the world I know or can know, and my knowledge is a physical part of this world.
  \end{enumerate}
  
  \item \textbf{Facts can be expressed as values of variables, which are ways systems interact.}
  \begin{enumerate}
    \item Properties are facts expressed by values \(a\) of a variable \(A\). The fact that \(A\) has value \(a\) is expressed as \(A=a\).
    \item Systems are collections of variables that admit approximate closed dynamical descriptions.
   \item Variables are ways systems interact and manifest themselves in interactions. 
        \item Physics is \emph{modal}: the general form of dynamical laws is: \emph{if \(a\) then \(b\)} with a certain reliability. Dynamical laws assign values to conditional probabilities \(p(b| a)\):  likelihood of \(b\) given \(a\).
 \item Causation is the aspect of modality that can support intervention.
  \item My world is coloured by my interaction with it.\footnote{\cite{Price}.} 
    
    \item Variables in a given perspective may have no value.\footnote{This aspect of the world was not known before the discovery of quantum phenomena.}
    \item In general, not all variables of a system can simultaneously have a value.    
        \item The mathematical theory allowing physical laws to be expressed associates elements \(A\) of a non-commutative algebra to the variables   \(A\)  of system.\footnote{This is not true if we disregard quantum phenomena, as Wittgenstein did.}
    \item Mathematical laws associated to a system relate variables — hence facts — relative to the same system, within the same perspective.
    \item A variable having a value in a perspective is an \emph{atomic fact}.      
    \item Combinations of systems are systems.
    \item Since variables express the way a system affects another in interactions, dynamical laws, which are about variables, regard interactions between systems.\footnote{Hence dynamical laws viewed as pertaining to a single system, such as a Schrödinger equation in isolation, are intrinsically incomplete.}
    \item The information about a system \(A\) in a perspective \(B\) changes when \(A\) and \(B\) interact. In the perspective \(B\), this is an update of information. In any other perspective, this is an evolution of the mutual information that is continuous in time.\footnote{The mis-named ``measurement problem" is this mismatch of perspectives.}
    \item The maximum information available about a system is achieved when a maximal set \(A\) of commuting variables has values that are facts.
    \item In a perspective \(B\), the facts making up the world are values of sets of variables \(A\). From a different perspective, the information that \(B\) has is embodied in  values of the commuting set of variables defining \(B\) and the fact that these are correlated to the \(A\)’s.
        \item I, as carrier of thought and knowledge, am a set of values of a set of variables that commute.\footnote{My personal knowledge is a physical configuration of my brain.}
      \end{enumerate}
    
  \item \textbf{Perspectives are transparent to each other because they are facts.}
  \begin{enumerate}
	\item An act of measurement is a special case of an ordinary physical interaction.
    \item A  system \(B\) can interact with a system \(A\) in many different manners. Accordingly, \(A\) may admit different layers of descriptions, all relative to \(B\). \label{layers}
      \item A perspective is accessible from another perspective, because it is simply the value of a family of (commuting) variables.
         \item The self is the reflective point of a knowledge that includes information about itself.\footnote{\cite{TIsmael2006-TISTSS} }
      \item Value attribution, such as \(A=a\), is a component of a language.
      \item The riddle of consciousness is the result of mistakenly assuming that a perspective is not a physical fact for any other perspective.
    \end{enumerate}
  
  \item \textbf{There is nothing wrong in circularity.}
  \begin{enumerate}
    \item Understanding perspective from other perspectives leads to circularity.
     \item Escaping circularity is not physical possibile, is logically incoherent and has no interest. It is the left over of an outdated metaphysics that limps in trying to make sense of new knowledge about the real world.
    \item The world of any perspective has a blind spot: the physical variables that embody this very perspective.
     \item Physics itself is in a perspective. 
    \item Within any perspective, circularity is broken indexically by that very perspective.\footnote{ \cite{TIsmael2006-TISTSS} }
    \item Information is always incomplete.
    \item The limits of my information are the limits of my world.
    \item The world is open.
    \item We can represent facts in a perspective with a proposition.
    \item All this is a useful conceptual structure compatible with current knowledge. Concepts change with learning. Any conceptual structure to frame reality is provisional.
  \end{enumerate}
  \item \textbf{What we do not have information about, we must pass over in silence.}
\end{enumerate}

\vspace{1cm}

\section*{Long}
\begin{enumerate}
  \item \textbf{The world is everything that is the case from some perspective.}
  \begin{enumerate}
    \item Everything that is the case, is the case from some perspective.
    \item Something that is the case from a perspective is a fact, from (or in) a perspective.
    \begin{enumerate}
      \item Each fact is a fact from a perspective.
      \begin{enumerate}
        \item The question whether there are facts outside a perspective has no meaning.
      \end{enumerate}
      \item A world is a collection of facts, not things.
      \item If you remove facts from a world, nothing remains.
    \end{enumerate}
    \item Our perspectives are special cases of the physical notion of perspective.\label{subject}
    \begin{enumerate}
      \item Perspectives are not specific to us, biological systems, societies, agents, computing systems, or similar.\label{nothuman}
      \item A perspective, in general, is not associated with mental or similar capacities.
      \item What is commonly called  ``an observer" is associated with a perspective, but not vice versa: the expression “observer” is commonly associated with restricted cases of a general perspective.
    \end{enumerate}
      \item Perspectives are not about minds. They are about physics, like reference systems in relativistic physics are about physics, not about minds.
       \begin{enumerate}
      	\item At this point, an attentive reader might be led to question the absoluteness of physical rules; this issue will be addressed in \ref{Physics},as we must climb the ladder one rung at a time in order to fully grasp the lesson proposed by physics. 
      \end{enumerate}
          \item No perspective stands out above the rest.
    \begin{enumerate}
      \item The reality of which we speak is the partial reality from within a perspective.
      \item The reality of which we speak can never be the total reality of all possible perspectives.\footnote{Cfr. \cite{Laozi2003-LAOTDO}, 1.1.}
      \begin{enumerate}
        \item Reality consists in states of affairs,\footnote{As expressed by \cite{Wittgensteintractatus} proposition 2.01.} but states of affairs do not exist outside a perspective.
        \item The question of whether a perspective-independent (“absolute”) reality exists, is a question without meaning.
      \end{enumerate}
    \end{enumerate}
  \end{enumerate}
  \item \textbf{My knowledge about the world is a particular kind of 
  perspective.}
    \begin{enumerate}
    \item My knowledge of the world  is based on information I have or can have about the world.
    \begin{enumerate}
      \item My (evolving) information about the world is my access to the world.
      \begin{enumerate}
        \item My perspective is my access to the world.
        \item My (evolving) information traces my perspective on the world.
      \end{enumerate}
      \item I can think of the world as empty, but this emptiness would still be counted as information I have about it.
      \item I can think about parts of the world I do not have information about, but their possibility is still information I have or can have.
    \end{enumerate}
    \item Information has degrees of reliability.\footnote{Cfr. \cite{De1989-FINPAC-2}.}
    \begin{enumerate}
      \item Degrees of reliability can be quantified: \(p(a) \in [0,1]\) quantifies my expectation that a possible fact \(a\) is going to be the case if I acquire information about it. \(p(a)=1\) indicates certainty that \(a\) is the case, and \(p(a)=0\) indicates certainty that \(a\) is not the case. 
    \end{enumerate}
    \item A fact in my perspective is knowledge about which I (reasonably) estimate no uncertainty.
    \begin{enumerate}
      \item A fact from my perspective is an \(a\) such that \(p(a) \approx 1\) and indicates (an almost) certainty that \(a\) is the case.
      \begin{enumerate}
        \item “There is a cup on the table I am looking at” is an example of such a fact.
        \item What I know with (sufficient) certainty constitutes a fact relative in my perspective.
      \end{enumerate}
    \end{enumerate}
    \item Degrees of reliability assigned to facts evolve; perspective can change accordingly.
    \begin{enumerate}
      \item \(a\) is a fact if I expect that new information will confirm it.
      \begin{enumerate}
        \item Once something is established as a fact, observing it again yields no additional information.
        \item A fact is what we call an element of physical reality.\footnote{Cfr. \cite{Einstein1935-EINCQD}, where an element of reality is defined in a very similar manner, as something I reliably expect to be true if observed. }
      \end{enumerate}
    \end{enumerate}
      \item To have uncertain information about a fact means to have the information that by gathering more information the fact will be ascertained with a known likelihood.
    \begin{enumerate}
    \item An \(a\) such that \(0 < p(a) < 1\) is an event about which I have information that lacks certainty.
    \end{enumerate}
    \item If I reliably expect that one out of \(N\) mutually exclusive alternatives \(a\) will be the case, then I can talk about a set of alternative possible facts.
    \begin{enumerate}   
    \item If reliably expect that one out of \(N\) mutually exclusive alternatives \(a\) to be the case, then  \(\sum_a p(a)=1.\)
      \item The uncertainty in the probability \(p(a)\) can be quantified by \(H(p) = -\sum_a p(a) \ln p(a)\), which is called the Shannon’s entropy of \(p(a)\).\footnote{\cite{Shannon}}
      \item If I know which of \(N\) alternatives is a fact, I have an information \(I(N) = \ln N,\) Shannon’s first definition of amount of information.\footnote{\cite{Shannon}}
      \item The amount of information in a distribution \(p(a)\) over \(N\) alternatives can be quantified by \(I(p) = I(N) - H(p).\) This vanishes if I do not know anything about which \(a\) is realized \((p(a)=constant)\), and is maximal (\(\ln N)\) if I know which \(a\) is realized (\(p(a)\) peaked on a single value).
      \begin{enumerate}
        \item \(p(a)\) expresses physical possibilities. No metaphysical possibility here, only physics.
      \end{enumerate}
    \end{enumerate}
    \item Facts in a perspective can be represented in logical spaces. They can be actual, or have probability.
    \begin{enumerate}
      \item Logical spaces are not the world. They are ways to represent perspectives.
      \item A logical space is the totality of all logically possible states of affairs.
      \begin{enumerate}
         \item  A logical space is a set of possible \(a\)'s and possible \(p(a)\)'s 
        \item What lies outside logical space is not false but meaningless. 
        \begin{enumerate}
        \item  To fall outside logical spaces is not negated, it is outside possibility.  
       \end{enumerate}
      \end{enumerate}
      \item The selection of a logical space depends on the perspective. No logical space is privileged.
      \item A fuzzy logical space\footnote{\cite{ZADEH1965338}} can account for the situation when we are unable to predicate facts with \(p(a)=1.\)
    \end{enumerate}
    \item My knowledge about the world does not create the world; it is determined by the world and it refers to it.
    \begin{enumerate}
      \item A perspective is not intended as a Dutch book.
      \begin{enumerate}
      \item Facts \(a\) and their probabilities \(p(a)\) form the physics of the world. They are well-defined also in the absence of credences or agents.
      \item  There is nothing in the physical explanation of the world to gamble on. 
      \end{enumerate}
    \end{enumerate}
  \end{enumerate}
  \item \textbf{Perspectives themselves are facts, when considered from another perspective.}
  \begin{enumerate}
    \item A perspective is embodied in a set of facts, when considered from any other perspective (including itself).
    \begin{enumerate}
    \item Here “Embodied” is not meant in the human or cognitive sense, but in a physical one: it applies to any physical object or system (as in \ref{nothuman}.)
           \item Perspectives are facts in two distinct senses: as the ensemble of actual facts {\em in} the perspective itself, and as the possible facts {\em embodying} the perspective, when seen in other perspectives.
    \end{enumerate}
    \item If knowing \(b\) determines which \(a\) is a fact, then \(a\) is fact with respect to \(b\), or in the perspective embodied in \(b\).
    \item If \(p(a,b)\) is the joint probability for two sets of possible facts \(A=\{a\}\) and \(B=\{b\},\) then \(b\) embodies a perspective in which the probabilities of \(a\) are \(p_b(a) = p(a|b) = p(a,b)/p(b).\)
    \begin{enumerate}
      \item \(a\) is a fact relative to \(b,\) or is a fact in the perspective \(b,\) if \(p_b(a)=1.\)
      \item Notice that \(a\) can be a fact relative to \(b\) (namely \(p_b(a)=1\)) whether or not it is a fact relative to me, to which the probabilities \(p(a,b)\) refer (that is, whether or not  \(p_a(a)=1\)). Indeed, \(a\) is a fact relative to \(b\) if it is a fact, but also if I have no certainty about \(a\) but I could get it by looking at \(B\). (Namely \(I_{A:B}\) is maximal, see \ref{mutual}.) 
    \end{enumerate}
    \item To know \textit{what} \(B\) knows about \(A\) is not the same as knowing \textit{that} \(B\) knows about \(A.\)
    \begin{enumerate}
      \item The second is a fact about \(B\) and the \(a\)’s; the first is an additional fact about \(B.\)
      \item The second is knowledge of a correlation; the first about its terms.
      \item If we can get information on a set of facts \(A=\{a\}\) by observing facts in a set \(B=\{b\},\) then \(B\) embodies a perspective on \(A.\)
      \begin{enumerate}
        \item The above is the case when \(A\) and \(B\) are correlated: knowledge of one informs about the other.
        \item If \(p(a,b)\) is the probability for the couple \((a,b)\) to be the case, then the marginals \(p_A(a)=\sum_b p(a,b)\) and \(p_B(b)=\sum_a p(a,b)\) are the respective probabilities for \(a\) to be the case and for \(b\)  to be the case, independently from one another; and \textit{the} amount of correlation in \(p(a,b)\) is measured by the mutual information \(I_{A:B} = I(p) - I(p_A) - I(p_B).\)
      \end{enumerate}
    \end{enumerate}
    \item The mutual information \(I_{A:B}\) quantifies the information on one variable that can be obtained from the other.\label{mutual}
    \begin{enumerate}
      \item By measuring \(b\), the average increase in information about \(a\), is given precisely by \(I_{A:B}.\)\footnote{\cite{RelativeinfoRovelli}}
      \item Mutual information quantifies the reduction in uncertainty about one random fact provided by knowledge of another.
      \begin{enumerate}
        \item Mutual information is a form of relative information.
      \end{enumerate}
    \end{enumerate}
 \item When the relative information \(I_{A|B}\)   is maximal, knowing \(b\) determines which \(a\) is a fact, and \(a\) is a fact relative to \(b\).\footnote{\cite{RelativeinfoRovelli}} \label{rfact}
    \begin{enumerate}
      \item This use of “fact” is consistent with 2.3 but refers to knowledge about the perspective \(B\) from a different perspective.
      \item A fact about a system in a perspective is accounted for, in any other perspective, either as a fact or, if not a fact in this other perspective, as relative information.
      \item What we are grasping here is the physical embodiment of perspectives, seen from other perspectives.
    \end{enumerate}
    \item All things stand only in relation to a perspective, and with a shift of perspective, what once appeared as fact may be absent or incompatible with other facts from another perspective.
    \begin{enumerate}
      \item All facts are relative facts.
    \end{enumerate}
    \item A subject is a (case of a) perspective and defines a world, as said in \ref{subject}. 
    \item From a perspective distinct from mine, I, a subject, am a physical process, namely a set of facts.
    \begin{enumerate}
      \item I, as carrier of thought and knowledge that instantiate a special case of perspective, am a set of facts both from my own perspective and in other perspectives. These facts concern my physical brain and its correlations with the else.
      \item Knowledge is physically embodied. There is no knowledge that is not embodied.
      \item My knowledge of the world is a set of facts regarding the physical process I am, and correlations with the rest of the world.\footnote{The emphasis on this point is novel with respect to \cite{Wittgensteintractatus}. Knowledge and language are themselves physical facts, they can be also understood as such, and doing so clarifies.}
    \end{enumerate}
    \item Knowledge is physically underpinned by relative information between variables of two systems: the subject and the object of knowledge.
    \begin{enumerate}
      \item Knowledge can appear as a perspective for the subject and as correlation from an external perspective.
      \item In physical interactions and more in general in time, knowledge  either  is acquired or lost.
      \begin{enumerate}
        \item Nothing different from physical interactions and time evolution happens in an agent or object of knowledge when knowledge is acquired or lost.
        \item Experience is a physical phenomenon, in which knowledge changes.
        \item Phenomenology is same, not distinct, from physics. The two are just different renderings of the same reality.
        \item The relation between experience and physical reality has been an enduring riddle in philosophy. Conceiving them as distinct a priori, rather than continuous, is the mistake: they are the same phenomenon seen in different perspectives. The answer to the riddle is that there is no riddle.
      \end{enumerate}
     \item In the classical limit, the concept of information acquires an epistemic dimension:  the observer’s --possibly partial-- knowledge about the objective perspective-independent state of affairs.
      \begin{enumerate}
      \item  The sharp distinction between facts and information about them makes only sense assuming absolute states of facts (see \ref{impossible}). 
      \item  In the quantum world, mutual information is always a measure of physical correlations. 
      \begin{enumerate}     
      \item The von Neumann's entropy of a density matrix \(\rho\) quantifies physical correlation.
            \begin{enumerate}
      \item By contrast, the measure introduced in \(3.10.3.2\) quantifies entanglement correlations between subsystems, rather than epistemic knowledge.        
       \end{enumerate}
      \end{enumerate}
       \item We can  recover of an epistemic notion of information in the sense of comparing --say-- limited information on a system available in a perspective, with richer information about the same system available in --say-- our own perspective. 
      \begin{enumerate}
      \item  For example when we quantify our knowledge of a quantum system, through the von Neumann entropy of the global state, we are adopting once again an epistemic notion of information.
      \end{enumerate}
      \end{enumerate}
      \item Meaning is evolutionarily — or biologically — relevant information.\footnote{\cite{Rovelli2018} and \cite{arcas2025intelligence}. }
      \begin{enumerate}
        \item Physics and biology together dispel the apparent magic in meaning.
        \item Physics clarifies the circular relation between knowledge as an epistemic concept and physical information in the physical world.
        \item The term “information” refers both to the gain in knowledge of a subject and to the correlations between systems. The latter is what we measure with relative information. The two are the same from different perspectives.
        \begin{enumerate}
        \item Information is not a primitive ontological ingredient; it is a structural feature of the correlations between systems.\footnote{This marks a departure from the information-theoretic interpretation proposed by \cite{Bub2010-BUBTDA-2}. }\color{black}
        \end{enumerate}

      \end{enumerate}
    \end{enumerate}
    \item The world for me is the world I know or can know, and my knowledge is part of this world.
    \begin{enumerate}
      \item There is nothing wrong in self-consistent circularity (see \(6\), below).
    \end{enumerate}
  \end{enumerate}
  
  \item \textbf{Facts can be expressed as values of variables, which are ways systems interact.}
  \begin{enumerate}
    \item Properties are facts expressed by values \(a\) of a variable \(A\). The fact that \(A\) has value \(a\) is expressed as \(A=a\).
    \item Systems are collections of variables that admit approximate closed dynamical descriptions (see \ref{dynamics}).
   \item Variables are ways systems interact and manifest themselves in interactions. 
  \begin{enumerate}
    \item Values of a variable \(A\) of a system label manners in which a different system can be affected in a certain interaction with the first system.
    \begin{enumerate}
    \item As facts are relative, so variables are the manners for a system to manifest itself in interacting with other systems.
    \item Measurements are cases of interactions leaving long-lasting traces, which increase the correlation between two systems.
      \begin{enumerate}
        \item These traces are not rare; they are not, like \textit{Dasein}\footnote{\citep{Heidegger1962-HEIBAT-8}. }, specific of humans. They are generic effects of physical interaction upon systems.
      \end{enumerate}
    \item The question “what is the value of a variable \(A\) of a system?” Formulated without (explicitly or implicitly) a perspective, is as meaningless as the question “what is the velocity of an object?” formulated without (explicitly or implicitly) a reference.
      \begin{enumerate}
        \item To ask about a property of a system apart from any perspective is to pose a question that has no meaning. Whatever is the case, is so from a perspective. Thinking otherwise is a misleading metaphysical assumption.
      \end{enumerate}
       \end{enumerate} 
       \end{enumerate} 
      \begin{enumerate}
        \item \label{dynamics} A dynamical description is a mathematical model of how values of variables change and are manifested. This is also called a “dynamical law”. 
        \item Physics is \emph{modal}: the general form of dynamical laws is: \emph{if \(a\) then \(b\)} with a certain reliability. That is, dynamical laws assign values to \(p(b| a)\):  likelihood of \(b\) given \(a\).
          \begin{enumerate}
            \item Physics is not  descriptive. It does not state or prescribes what \emph{is} the case. It states what \emph{may} be the case if something else is the case. It restricts the space of possibilities; it does not fix one single possibility.
            \item Modality is not a signal of antirealism. It is the expression of the regularities we have found in the real world.
            \end{enumerate}
      \end{enumerate}

    \item Variables in a given perspective may have no value.\footnote{This aspect of the world was not known before the discovery of quantum phenomena.}
      \begin{enumerate}
        \item If a variable \(A\) has no value and \(p(a)+p(a')=1\), then it may be that there is a \(c\) such that
        \(p(c\mid a)p(a)+p(c\mid a')p(a')\neq p(c)\). This is called \emph{interference}.\footnote{This characterizes quantum phenomena (quantum interference).}
          \begin{enumerate}
            \item The above is not a contradiction, because it may be impossible to ascertain \(p(c)\) and \(p(a)\) independently (see below). Hence \(p(a)\) and \(p(c)\) refer to two different situations. 
        \item In this case, we say that \(a\) and \(a'\) are “in quantum superposition”.
        
          \begin{enumerate}
            \item The term comes from quantum phenomena, where it is the case that a system may exist in a superposition of states.
          \end{enumerate}
          \end{enumerate}
        \item \(A=a\) can be a fact in a perspective, and not be a fact in another perspective.
          \begin{enumerate}
            \item \(A=a\) can be a fact in our perspective and not be a fact in another perspective.
              \begin{enumerate}
                \item The possibility in 4.3.2.1 is realized, for instance, if in the other perspective we are outselves a term “in a quantum superposition”.
                \item If \(a\) is a term in a quantum superposition in a perspective, \(a\) is not a fact in that perspective.
                \item The point \color{red} 4.3.2.2. \color{black} is what forbids us to take facts that we observe as absolute, beyond our perspective. These are the limits of naïve realism.
              \end{enumerate}
          \end{enumerate}
      \end{enumerate}

    \item In general, not all variables of a system can simultaneously have a value.      \begin{enumerate}
        \item Two variables that can always have values together are said to \emph{commute}.
          \begin{enumerate}
            \item Systems generically have variables that do not commute.
            \item If two variables \(A\) and \(B\) commute, we can always obtain new information about \(A\) without affecting the information we have about \(B\). Not so if they do not commute, because in this case since they may not have values together, getting information about one can render the information we had about the other unreliable.\footnote{Heisenberg’s uncertainty.}
          \end{enumerate}
      \end{enumerate}

    \item The mathematical theory allowing physical laws to be expressed associates elements \(A\) of a non-commutative algebra to the variables  \(A\)  of system.\footnote{This is not true if we disregard quantum phenomena, as Wittgenstein did.}
      \begin{enumerate}
        \item A non-commutative algebra is a set where addition and multiplication are defined but, in general, \(AB\neq BA\). This reflects the fact that, in getting information about \(A\) and \(B\), the order in which we do so matters.
        \item The eigenvalues of an algebra element \(A\) are the possible values \(a\) that the variable \(A\) can take.\footnote{The eigenvalues of an algebra element \(A\) are the real numbers \(a\) for which \((A-a\,\mathbf{1})\) has no inverse.}
        \item The facts that embody a perspective are represented by simultaneous values of  variables. Therefore a perspective is defined by an \emph{abelian subalgebra} of the algebra of variables. An abelian subalgebra is a linear subset of an algebra formed by elements that commute (such that \(AB=BA\)).\footnote{\cite{RelativeinfoRovelli}.}
          \begin{enumerate}
            \item Information is carried by values of abelian subalgebras.
            \item A perspective is given by the facts and the probabilities relative to the values of the variables of an abelian subalgebra of a system.
          \end{enumerate}
        \item In a given perspective, the “state” of another system is the information about that system available in that perspective.
        \item The state of a system relative to a perspective is given by a \emph{linear} functional \(\rho\) (a “quantum state”) on the algebra of the variables of this system. Given the spectral decomposition \(A=\sum_a a\,\Pi_a\), of a variable \(A\), the probability \(p(a)\) is given by \(p(a)=\rho(\Pi_a)\). For two commuting variables \(A,B\), \(p(a,b)=\rho(\Pi_a \Pi_b)\). This shows that the quantum formalism includes the precise mathematical tools for defining  relative facts, as detailed in \ref{rfact}.
      \end{enumerate}

    \item Mathematical laws associated to a system relate variables — hence facts — relative to the same system, within the same perspective.
      \begin{enumerate}
        \item Mathematical laws limit how values of variables can change together.
          \begin{enumerate}
            \item They make the world partially predictable and intelligible.
          \end{enumerate}
        \item A world has a logical form made by a family of possible relative facts \(a,b,\ldots\) and likelihoods \(p(a)\) given by conditionals \(p(a| b)\). This allows any sufficiently competent agent to have expectations about what new knowledge could bring. \label{possible}
      \end{enumerate}

    \item A variable having a value in a perspective is an \emph{atomic fact}.
      \begin{enumerate}
        \item An atomic fact is the simplest entity.
        \item An atomic fact involves always a variable and a perspective.
          \begin{enumerate}
            \item This is the most fundamental way to have an atomic fact; nothing can be made in a less structured way.
            \item We say that systems and in particular objects (which are bundles of properties\footnote{\citep{Simons1994-SIMPIP}.}) exist, insofar as they are manifest in a perspective (a relation). This is a decisive break from the older conception of things.\footnote{See \cite{Wittgensteintractatus}, \S3.25.}
          \end{enumerate}
        \item An atomic fact is the possibility of the value of a variable occurring in a perspective.\footnote{See \cite{Wittgensteintractatus},  \S2.0123.}
        \item An atomic fact \(a\) is accounted for as an eigenvalue of an algebra element \(A\).
          \begin{enumerate}
            \item The totality of actual atomic facts in a perspective is the world from that perspective.
              \begin{enumerate}
                \item We can distinguish two kinds of atomic facts, following the lesson learned from the Master.
                \item  \emph{Sachverhalte} are the atomic facts as defined in \ref{possible}, the possible configurations within logical space. 
                \item  \emph{Tatsachen} are the \textit{actual atomic facts} that make the world always a set of facts that are the case (see \ref{fact})
              \end{enumerate}
            \item All perspectives are themselves in a space made of possible atomic facts. I can think of this space as empty, but not of a perspective without this space.
            \item It would then be impossible to draw up a picture of the world without a perspective \label{impossible}.\footnote{See \cite{Wittgensteintractatus}, \S 2.0212.}
          \end{enumerate}
      
          \end{enumerate}

    \item Combinations of systems are systems.
      \begin{enumerate}
        \item Facts regarding combined systems may sometimes be determined by facts regarding the individual systems that compose them. But not always so.\footnote{Failure of this is called entanglement. It does not happen if we disregard quantum phenomena.}
      \end{enumerate}
    \item Since variables express the way a system affects another in interactions, dynamical laws, which are about variables, regard interactions between systems.\footnote{Hence dynamical laws viewed as pertaining to a single system, such as a Schrödinger equation in isolation, are intrinsically incomplete.}
      \begin{enumerate}
        \item Dynamics regards relations.
        \item In the non-relativistic limit, dynamical laws can be expressed as evolution in a preferred time variable.
      \end{enumerate}
    \item The information about a system \(A\) in a perspective \(B\) changes when \(A\) and \(B\) interact. In the perspective \(B\), this is an update of information. In any other perspective, this is an evolution of the mutual information that is continuous in time.\footnote{The mis-named "measurement problem" is this mismatch of perspectives.}

    \item The maximum information available about a system is achieved when a maximal set \(A\) of commuting variables has values that are facts.
      \begin{enumerate}
        \item When this is the case, Shannon entropy is minimal.
        \item The maximum information available about a system is insufficient to determine the further information that the system can provide, because values of variables that do not commute with the known ones are not facts. Interactions with these variables bring about genuinely new information. This does not increase the maximum information about the system because it renders previous information irrelevant: some facts cease to be facts.\footnote{The two postulates in axiomatic formulation of RQM, \cite{Rovelli1996-ROVRQM} are (i) system’s relevant information is limited, but (ii) new information can always be acquired.}
          \begin{enumerate}
            \item A fact ceases to be a fact once its probability \(p(a)\) becomes less than one. This can happen after a new interaction, or simply as time passes.
          \end{enumerate}
        \item A perspective may be understood as the information concerning facts relative to a system, which in a particular can be (but needs not to be) a complex observer.
      \end{enumerate}

    \item In a perspective \(B\), the facts making up the world are values of sets of variables \(A\). From a different perspective, the information that \(B\) has is embodied in  values of the commuting set of variables defining \(B\) and the fact that these are correlated to the \(A\)’s.
        \item I, as carrier of thought and knowledge, am a set of values of a set of variables that commute.\footnote{My personal knowledge is a physical configuration of my brain.}
      \end{enumerate}

  \item \textbf{Perspectives are transparent to one another because they are facts.}
  \begin{enumerate}
	\item An act of measurement is a special case of an ordinary physical interaction.
	\begin{enumerate}
		\item What I know or can know about the events \(a\) and \(b\) is given by their joint probability \(p(a,b)\). From this, the perspective of \(b\) on \(a\) is defined by the conditional probabilities \(p(a|b)\).
		\item Knowledge, as stated in \ref{knowledge} is physically underpinned by a physical configuration correlated with the part of the world it concerns. Because it is physical, it is accessible from other perspectives through interaction. No perspective lies beyond the physical.
	\end{enumerate} 
    \item A  system \(B\) can interact with a system \(A\) in many different manners. Accordingly, \(A\) may admit different layers of descriptions, all relative to \(B\).
      \begin{enumerate}
        \item The same subject can have information about the same system organized in different layers.
        \item Layers in the world are approximate descriptions of the same set of facts\footnote{For instance: physical, chemical, biological, functional, psychological, social, \dots} reflecting different modes of physical interaction.
          \begin{enumerate}
            \item For instance, “The forest has grown” and “New trees are born near the old ones” refer to two layers of description of the same events.
            \item Different layers define different logical spaces and may, though not necessarily, require different logics.
          \end{enumerate}
        \item The world admits different layers of description.
        \item Layers are distinct and have diverse efficacy, but are never in contradiction. Nature is coherent.
      \end{enumerate} 
    \item A perspective can be accessible from another perspective, because it is simply (embodied in) the value of a family of (commuting) variables.
       \item My perspective is my world.
      \begin{enumerate}
        \item Humans are systems that embody perspectives, but are peculiar on many grounds.
        \item My own perspective is a limit of my world. The variables where its information is stored are accessible by any other perspective.
       \end{enumerate}
        \item The self is the reflective point of a knowledge that includes information about itself.\footnote{\cite{TIsmael2006-TISTSS} }
          \begin{enumerate}
            \item Knowledge about ourselves is always vastly incomplete.\footnote{\citep{Spinoza2020-SPISE}.}   
        \item Values of variables, including those forming a perspective, hence knowledge, hence the self, are physical phenomena.
        \item Perspectives are physical facts: there is no fact that is not also a physical fact. Yet, most phenomena are far better described using the language of special layers, because as physical facts they are obscure. Appropriate language makes them clear and understandable.
          \begin{enumerate}
            \item Thought is a physical phenomenon; it can in principle be described as a set of facts.\footnote{\citep{Spinoza2020-SPISE}. This is also what neurosciences aim to do. } Ethics is a physical phenomenon; it can be also be in principle described as a set of facts. But they are both phenomena better described in appropriate layers as in \ref{layers}.
              \begin{enumerate}
                \item That each layer admits its own logical space does not divide the world into distinct phenomena.
              \end{enumerate}
            \item The transparency of a perspective to another does not need to be realized at the physical level. It works more easily at higher levels, with the associated uncertainty. This is called “communicating”.\footnote{\cite{Zhuangzi}, 17.}
          \end{enumerate}
      \end{enumerate}
    \item Value attribution, such as \(A=a\), is a component of a language.
      \begin{enumerate}
        \item A language can express the knowledge in a perspective.
        \item A language is a set of enunciations.
          \begin{enumerate}
            \item Enunciations are propositions expressed according to the rules of the logic that underlies the chosen perspective.
          \end{enumerate}
        \item An enunciation in a language is itself a fact in a world. It is not outside it.
        \item No enunciation exists unless it is enunciated.
          \begin{enumerate}
            \item There is no un-embodied language.\footnote{This is the the central difference from \cite{Wittgensteintractatus}. }
            \item A world lives in a logical space, and any logical proposition is a fact in a world.
          \end{enumerate}
        \item A proposition shows the logical form of a perspective.
          \begin{enumerate}
            \item No unique logical form can be assumed; the form of logic itself depends on the layer under consideration.
          \end{enumerate}
        \item The general form of physical knowledge can be expressed both in laws of the form \(p(a|b)\) and in the structure of logical spaces.
      \end{enumerate}
    \item The riddle of consciousness is the result of mistakenly assuming that a perspective is not a physical fact for any other perspective.
      \begin{enumerate}
        \item The hard problem of consciousness stems from assuming, not from observing, that there are facts that are not physical facts. All facts are simply facts in a perspective.
          \begin{enumerate}
            \item There are no physical facts outside a perspective, nor perspectives which are not physical facts.
            \item It is not the first-person perspective that is mysterious. It is the absolute third-person perspective that is non-existing.
          \end{enumerate}
      \end{enumerate}
  \end{enumerate}
  
  \item \textbf{There is nothing wrong in circularity.}
  \begin{enumerate}
    \item Understanding perspective from other perspectives leads to circularity (as in \ref{3}, above).
      \begin{enumerate}
        \item Circularity in understanding occurs when some phenomenon is accounted for by assuming the phenomenon itself (perspectives can be accounted for from other perspectives).
        \item Searching for an escape from circularity by regressing to larger and larger perspectives leads to an infinite regress.
        \item Escaping circularity is a request for more than relative information: it is a search for un-embodied information.    But the idea of un-embodied information is a metaphysical dream (dreamed by embodied creatures).
      \end{enumerate}
    \item Escaping circularity is not physical possibile, is logically incoherent and has no interest. It is the left over of an outdated metaphysics that limps in trying to make sense of new knowledge about the real world.
    \item The world of any perspective has a blind spot: the physical variables that embody this very perspective.
      \begin{enumerate}
        \item These same variables can be accounted for from another perspective. But there is no way to account for all perspectives, because of \(1.4.2\), \ref{knowledge} and \(5.4.\)   A world is only such with respect to something physical, when quantum phenomena are not disregarded.  This gives rise to the circularity of making sense of perspective, from other perspectives, as correlations. 
        \item Circularity appears to be a problem in logical analysis only when trying to make claims outside any perspective. Such claims cannot be made by any embodied speaker.
          \begin{enumerate}
            \item On logical analysis, circularity is already employed in set theory through the theorem stating that “every graph has a unique decoration.”\footnote{\cite{Aczel1988-ACZNS} chapter \(6\).} In this way, it becomes possible to define a graph in terms of itself, something forbidden in Zermelo–Frankel set theory. This is useful circularity.
          \end{enumerate}
      \end{enumerate}

      \item Physics itself is in a perspective. \label{Physics}
      \begin{enumerate}
      	\item  \ In physics, we use variables, functions and rules, and mistakenly consider these absolute. This gives rise to the wrong idea of a series of absolute rules outside any perspective. 
      	But the rules, in fact, are part of our knowledge, and as such perspectival themselves. 
      \end{enumerate}

    \item Within any perspective, circularity is broken indexically by that very perspective.
      \begin{enumerate}
        \item The idea of a single fundamental perspective from which everything would flow is flawed.
        \item It is the idea of a universal foundation of reality that is flawed.
      \end{enumerate}
  \item Information is always incomplete
      \begin{enumerate}
        \item It is always possible for a system to acquire new information.\footnote{This is the second postulate in axiomatic formulation of RQM.}
        		   \begin{enumerate}
		  	 \item Quantum non-predictability renders information permanently incomplete. 
                           \item Interactions between systems change correlations and mutual information rendering information widespread and ever-changing. 
           		 \item The emergence of new correlations yields new information and changing perspectives.
        	\item The world cannot be described from a closed standpoint. 
           		 \item It is never the case that a new interaction fails to yield information. When we speak of “no new information,” this is only because we have forgotten that the observer’s knowledge is itself a perspective.
          		\end{enumerate}
      	  \item Information is always partial because it is incomplete and perspectival. 
	         \end{enumerate} 
         \item The limits of my information are the limits of my world.
   		\begin{enumerate} 
		\item The limits of my world move continuously, because I learn and my information changes.
		       \item  My knowledge is always incomplete.
		  \end{enumerate} 
        \item The world is open.
          \begin{enumerate}
          \item My world is my perspective. 
        \item We know with confidence that reality is wider than the information we have about it. Lack of information does not imply knowledge of absence. Nor does it implies  knowledge about what is not known.
            \item To think that we can have a picture of all facts from all possible perspectives is childish.
                \item Our ignorance of what we do not know and could learn is not lack of knowledge in the sense of Shannon’s theory, because we do not have knowledge of a fixed space of possibilities for what we don’t know.
       \end{enumerate}
    \item We can represent facts in a perspective with a proposition.
      \begin{enumerate}
        \item A set of propositions is a model.
        \item A model is not reality. It is a shard of reality.
          	\begin{enumerate}
           	 \item Looking through a model is to see reality from a point of view, like from a part of a mirror.
              		\begin{enumerate}
                		\item The mirror is never whole.
                		\item It is not ignorance that makes the mirror incomplete, but the perspectival nature of reality.
              		\end{enumerate}
          	\end{enumerate}
      	\item  The idea of the ensemble formed by the totality of all perspectives, and of the facts within them, does not belong to reality, for it is not contained within any perspective.\footnote{The emphasis on this point is connected to \cite{Wittgensteintractatus} (Proposition 5.6), where the author reflects on the limits of language: the boundary between what can be said and what can only be shown.}
          	\begin{enumerate}
      		\item Quantum phenomena reveal what should have been evident all along: since any knower is part of reality, all knowledge is necessarily perspectival. 
            	\item Quantum phenomena only block a wrong metaphysical jump that is not granted by anything but seemed to allowed by classical physics: the idea of accessing a completely perspective independent reality. 
          	\end{enumerate}
        \item By reality we may indicate the content of any perspective, or the idea of all perspectives, because these can be the case from a perspective, but not the collection of all perspectives, which is not something that is the case from any perspective.
        \item The totality of the perspectives cannot be a perspective because it lacks embodiment.
        \item The ghost of an absolute unaccessible reality is useless, because it plays no role for us except for troubling us for nothing.
      		\end{enumerate} 
     \item All this is a useful conceptual structure compatible with current knowledge. Concepts change with learning. Any conceptual structure to frame reality is provisional.
      		\begin{enumerate}
       		\item Hence all of the above is just a perspective on perspectives, and as such partial. It is not a fundamental view.\footnote{\citep{Ngrjuna} \(MMK.13.8\). }
        		\item Anything posed as ultimately fundamental, or ultimately primary, is uninteresting.
   \end{enumerate}
  \end{enumerate} 

  \item \textbf{What we do not have information about, we must pass over in silence.}
\end{enumerate}

\vspace{2cm}

\bibliographystyle{chicago}

\bibliography{Bibliography}

\begin{thebibliography}{}

\bibitem[\protect\citeauthoryear{Aczel}{Aczel}{1988}]{Aczel1988-ACZNS}
Aczel, P. (1988).
\newblock {\em Non-Well-Founded Sets}.
\newblock Palo Alto, CA, USA: Csli Lecture Notes.

\bibitem[\protect\citeauthoryear{Arcas}{Arcas}{2025}]{arcas2025intelligence}
Arcas, B. (2025).
\newblock {\em What Is Intelligence?: Lessons from AI About Evolution,
  Computing, and Minds}.
\newblock Antikythera. MIT Press.

\bibitem[\protect\citeauthoryear{Bub and Pitowsky}{Bub and
  Pitowsky}{2010}]{Bub2010-BUBTDA-2}
Bub, J. and I.~Pitowsky (2010).
\newblock Two dogmas about quantum mechanics.
\newblock In S.~Saunders, J.~Barrett, A.~Kent, and D.~Wallace (Eds.), {\em Many
  Worlds?: Everett, Quantum Theory \& Reality}. Oxford University Press UK.

\bibitem[\protect\citeauthoryear{{De Finetti}}{{De
  Finetti}}{1989}]{De1989-FINPAC-2}
{De Finetti}, B. (1989).
\newblock Probabilism: A critical essay on the theory of probability and on the
  value of science.
\newblock {\em Erkenntnis\/}~{\em 31\/}(2-3), 169--223.

\bibitem[\protect\citeauthoryear{{Di Biagio} and Rovelli}{{Di Biagio} and
  Rovelli}{2025}]{RelativeinfoRovelli}
{Di Biagio}, A. and C.~Rovelli (2025, October).
\newblock Relative information, relative facts.

\bibitem[\protect\citeauthoryear{Einstein, Podolsky, and Rosen}{Einstein
  et~al.}{1935}]{Einstein1935-EINCQD}
Einstein, A., B.~Podolsky, and N.~Rosen (1935).
\newblock Can quantum-mechanical description of physical reality be considered
  complete?
\newblock {\em Physical Review\/}~(47), 777--780.

\bibitem[\protect\citeauthoryear{Heidegger}{Heidegger}{1962}]{Heidegger1962-HEIBAT-8}
Heidegger, M. (1962).
\newblock {\em Being and Time}.
\newblock New York,: Harper.

\bibitem[\protect\citeauthoryear{Ismael}{Ismael}{2006}]{TIsmael2006-TISTSS}
Ismael, J.~T. (2006).
\newblock {\em The Situated Self}.
\newblock New York, US: Oxford University Press USA.

\bibitem[\protect\citeauthoryear{Laozi and Ivanhoe}{Laozi and
  Ivanhoe}{2003}]{Laozi2003-LAOTDO}
Laozi and P.~J. Ivanhoe (2003).
\newblock {\em The Daodejing of Laozi}.
\newblock Hackett Publishing Company.

\bibitem[\protect\citeauthoryear{Price}{Price}{2007}]{Price}
Price, H. (2007).
\newblock Causal perspectivalism.
\newblock In H.~Price and R.~Corry (Eds.), {\em Causation, Physics, and the
  Constitution of Reality: Russell's Republic Revisited}, pp.\  250--292.
  Oxford: Oxford University Press.

\bibitem[\protect\citeauthoryear{Rovelli}{Rovelli}{1996}]{Rovelli1996-ROVRQM}
Rovelli, C. (1996).
\newblock Relational quantum mechanics.
\newblock {\em International Journal of Theoretical Physics\/}~{\em 35\/}(8),
  1637--1678.

\bibitem[\protect\citeauthoryear{Rovelli}{Rovelli}{2018}]{Rovelli2018}
Rovelli, C. (2018).
\newblock {\em Meaning and Intentionality = Information + Evolution}, pp.\
  17--27.
\newblock Cham: Springer International Publishing.

\bibitem[\protect\citeauthoryear{Shannon}{Shannon}{1948}]{Shannon}
Shannon, C.~E. (1948).
\newblock A mathematical theory of communication.
\newblock {\em The Bell System Technical Journal\/}~{\em 27\/}(3), 379--423.

\bibitem[\protect\citeauthoryear{Siderits}{Siderits}{2022}]{Ngrjuna}
Siderits, M. (2022, 05).
\newblock Nāgārjuna.

\bibitem[\protect\citeauthoryear{Simons}{Simons}{1994}]{Simons1994-SIMPIP}
Simons, P. (1994).
\newblock Particulars in particular clothing: Three trope theories of
  substance.
\newblock {\em Philosophy and Phenomenological Research\/}~{\em 54\/}(3),
  553--575.

\bibitem[\protect\citeauthoryear{Spinoza}{Spinoza}{2020}]{Spinoza2020-SPISE}
Spinoza, B. (2020).
\newblock {\em Spinoza's Ethics}.
\newblock Princeton: Princeton University Press.

\bibitem[\protect\citeauthoryear{Watson}{Watson}{2003}]{Zhuangzi}
Watson, B. (2003).
\newblock {\em Zhuangzi: Basic Writings}.
\newblock Columbia University Press.

\bibitem[\protect\citeauthoryear{Wittgenstein}{Wittgenstein}{1922}]{Wittgensteintractatus}
Wittgenstein, L. (1922).
\newblock {\em Tractatus Logico-Philosophicus}.
\newblock London: Routledge \& Kegan Paul.
\newblock Originally published as ``Logisch-Philosophische Abhandlung'' in
  \textit{Annalen der Naturphilosophie}, XIV (3/4), 1921.

\bibitem[\protect\citeauthoryear{Wittgenstein}{Wittgenstein}{1953}]{WittgensteinPI}
Wittgenstein, L. (1953).
\newblock {\em Philosophical Investigations}.
\newblock New York, NY, USA: Wiley-Blackwell.

\bibitem[\protect\citeauthoryear{Zadeh}{Zadeh}{1965}]{ZADEH1965338}
Zadeh, L. (1965).
\newblock Fuzzy sets.
\newblock {\em Information and Control\/}~{\em 8\/}(3), 338--353.

\end{thebibliography}
\end{document}